\documentclass[useAMS,usenatbib]{mn2e}
\usepackage{txfonts}
\usepackage{graphicx}
\usepackage{natbib}
\usepackage{if then}

\def\del#1{{}}

\newcommand{\unit}[1]{\mbox{  }\rm{#1}}

\newcommand{\reack}[1]
{
\ifthenelse{\equal{#1}{k1}}{\(\rm H+e^{-}\to H^{+}+2e^{-}\)}
{\ifthenelse{\equal{#1}{k2}}{\(\rm H^{+}+e^{-}\to H+\gamma\)}{\ifthenelse{\equal{#1}{k3}}{\(\rm He+e^{-}\to He^{+}+2e^{-}\)}{\ifthenelse{\equal{#1}{k4}}{\(\rm He^{+}+e^{-}\to He+\gamma\)}{\ifthenelse{\equal{#1}{k5}}{\(\rm He^{+}+e^{-}\to He^{++}+2e^{-}\)}{\ifthenelse{\equal{#1}{k6}}{\(\rm He^{++}+e^{-}\to He^{+}+\gamma\)}{\ifthenelse{\equal{#1}{k7}}{\(\rm H+e^{-}\to H^{-}+\gamma\)}{\ifthenelse{\equal{#1}{k8}}{\(\rm H+H^{-}\to H_{2}+e^{-}\)}{\ifthenelse{\equal{#1}{k9}}{\(\rm H+H^{+}\to H_{2}^{+}+\gamma\)}{\ifthenelse{\equal{#1}{k10}}{\(\rm H_{2}^{+}+H\to H_{2}+H^{+}\)}{\ifthenelse{\equal{#1}{k11}}{\(\rm H_{2}+H^{+}\to H_{2}^{+}+H\)}{\ifthenelse{\equal{#1}{k12}}{\(\rm H_{2}+e^{-}\to 2H+e^{-}\)}{\ifthenelse{\equal{#1}{k13}}{\(\rm H_{2}+H\to 3H\)}{\ifthenelse{\equal{#1}{k14}}{\(\rm H^{-}+e^{-}\to H+2e^{-}\)}{\ifthenelse{\equal{#1}{k15}}{\(\rm H^{-}+H\to 2H+e^{-}\)}{\ifthenelse{\equal{#1}{k16}}{\(\rm H^{-}+H^{+}\to 2H\)}{\ifthenelse{\equal{#1}{k17}}{\(\rm H^{-}+H^{+}\to H_{2}^{+}+e^{-}\)}{\ifthenelse{\equal{#1}{k18}}{\(\rm H_{2}^{+}+e^{-}\to 2H\)}{\ifthenelse{\equal{#1}{k19}}{\(\rm H_{2}^{+}+H^{-}\to H_{2}+H\)}{\ifthenelse{\equal{#1}{k29}}{\(\rm D^{+}+e^{-}\to D+\gamma\)}{\ifthenelse{\equal{#1}{k31}}{\(\rm D+H^{+}\to D^{+}+H\)}{\ifthenelse{\equal{#1}{k32}}{\(\rm D^{+}+H\to D+H^{+}\)}{\ifthenelse{\equal{#1}{k33}}{\(\rm D^{+}+H_{2}\to HD+H^{+}\)}{\ifthenelse{\equal{#1}{k34}}{\(\rm HD+H^{+}\to H_{2}+D^{+}\)}{\ifthenelse{\equal{#1}{k50}}{\(\rm He+H^{+}\to HeH^{+}+\gamma\)}{\ifthenelse{\equal{#1}{k51}}{\(\rm HeH^{+}+H\to He+H_{2}^{+}\)}{\ifthenelse{\equal{#1}{k52}}{\(\rm HeH^{+}+e^{-}\to He+H\)}{
}}}}}}}}}}}}}}}}}}}}}}}}}}}}
\title{Cosmic rays and the primordial gas}
\author[Jens Jasche \\$^{1}$, Benedetta Ciardi \& Torsten A. En\ss lin]
       {Jens Jasche $^{1}$, Benedetta Ciardi $^{1}$, Torsten A. En\ss lin $^{1}$ \\$^{1}$ Max-Planck-Institut f\"{u}r Astrophysik , Karl-Schwarzschild Strasse 1,  D-85748 Garching, Germany\\}
\begin{document}
\date{Submitted to MNRAS 03-Mar-2007}

\pagerange{\pageref{firstpage}--\pageref{lastpage}} \pubyear{2006}

\maketitle

\label{firstpage}
\begin{abstract}
One of the most outstanding problems in the gravitational collapse scenario of early structure formation is the cooling of primordial gas to allow for small mass objects to form. As the neutral primordial gas is a poor radiator at temperatures \(T\le10^4\unit{K}\), molecular hydrogen is needed for further cooling down to temperatures \(T\sim100\unit{K}\). The formation of molecular hydrogen is catalyzed by the presence of free electrons, which could be provided by the ionization due to an early population of cosmic rays. In order to investigate this possibility we developed a code to study the effects of ionizing  cosmic rays  on the thermal and chemical evolution of primordial gas. We found that cosmic rays can provide enough free electrons needed for the formation of molecular hydrogen, and therefore can increase the cooling ability of such primordial gas under following conditions: A dissociating photon flux with \(F<10^{-18}\unit{erg\,cm^{-2}\,Hz^{-1}\,s^{-1}}\), initial temperature of the gas \(\sim10^{3}\unit{K}\), total gas number densities \(n\ge1\unit{cm^{-3}}\), Cosmic ray sources with \(\dot{\epsilon}_{CR}>10^{-33}\unit{erg\,cm^{-3}\,s^{-1}}\).
\end{abstract}


\section{Introduction}

In the commonly adopted theory for structure formation, primordial density
perturbations grow through gravitational Jeans' instability and
eventually form halos in whose potential wells the observable baryonic matter gathers. 
During contraction, the gas develops shocks and gets reheated to a
temperature at which pressure support can prevent further collapse. Once the gas has virialized in 
the potential wells of pre-existing dark matter halos, additional
cooling is then required to further collapse the gas and form luminous objects.

Line cooling, in which the radiative de-excitation of atoms leads to
the emission of photons, is the most important cooling mechanism for a
primordial gas (predominantly made of hydrogen and helium) at temperatures $T > 10^4$~K.
At lower temperatures though, primordial gas is a very poor radiator
\citep{shapiro_kang} and in the absence of elements with lower excitation energies, such as metals, 
line cooling becomes ineffective \citep{Galli_Palla}. In the standard 
cosmological hierarchical scenario for structure formation the objects
which form first are predicted to have masses corresponding to virial 
temperatures \(T_{vir}<10^4\unit{K}\) \citep[for a review see][]{Ciardi}. Thus these objects, 
once virialized, cannot cool and further collapse via atomic line cooling.

The most efficient coolant for a primordial gas at such temperatures
is molecular hydrogen, because it has additional rotational and
vibrational degrees of freedom and therefore is able to cool the gas
down to temperatures \(T\sim100\unit{K}\). For this reason molecular
hydrogen is believed to play an important role in the formation of
stars or small-mass galaxies from a metal-free gas. 

Therefore, it is of great interest to get a firm picture of the primordial chemistry and of all the effects that might lead to formation or destruction of molecular hydrogen in metal-free gas.
At redshifts \(z\le 110\) molecular hydrogen is predominantly formed by the gas phase reactions
\begin{eqnarray}
\label{eqn:h2channel}
\rm{H}+\rm{e^-} & \to & \rm{H^-}+ \gamma\nonumber \\
\rm{H^-}+\rm{H} & \to & \rm{H_2}+ \rm{e^-}\nonumber 
\end{eqnarray}
in which electrons act as a catalyst \citep{dalgarno_lepp}. Thus, the
formation of molecular hydrogen requires free electrons to be abundant
at relatively low temperatures (\(T\le10^4\unit{K}\)). 

In a post
recombination universe the gas is mainly neutral with a residual
fraction of free electrons which give rise to the formation of traces
of molecular hydrogen \citep{Galli_Palla}. Nevertheless, this
primordial abundance is not enough to trigger
structure formation \citep{Hirata,Abel_chem,Tegmark,Haimann}. Partial ionization of hydrogen and
helium increases the abundance of free electrons and
promotes the formation of molecular hydrogen. While UV photons can
fully ionize such species and, in the absence of neutral hydrogen,
prevent the formation of H$_2$, x-rays only produce a partial
ionization and promote molecular hydrogen formation. In the past,
several authors have investigated the influence of x-rays on the H$_2$
chemistry (e.g. \citet{Glover,Haiman_2000}).

Recently much work has been published on the effect on primordial chemistry of high
energetic cosmic rays arising from the decay of massive dark matter
particles \citep[e.g.][]{Shchekinov,Ripamonti}. \citet{Shchekinov}
pointed out that ultra high energetic cosmic rays (UHECR), arising
from decaying super heavy dark matter particles with masses
\(M_X\ge10^{12}\unit{GeV}\) may have a strong effect on the chemical
evolution of the medium. Those high energy cosmic rays are believed to
be converted into UV photons through electromagnetic cascades which
would partially ionize hydrogen and eventually lead to
the formation of molecular hydrogen. The authors showed that in the
presence of UHECR the early luminous objects are an order of magnitude
less massive and form earlier in comparison to the predictions of the
standard recombination history.

Cosmic rays (CR) themselves can be an alternative source of partial
ionization. Unlike photons, cosmic ray particles can ionize neutral atoms many
times as they move through the medium. In addition, in each ionizing collision the
cosmic ray particles impart only a small fraction of their overall
kinetic energy to the ejected electron
\citep{Mannheim_Schlickeiser,Spitzer_Scott}. Therefore cosmic rays are
expected to heat the medium much less than photons with comparable
energies.
There are several suggestions about possible cosmic ray sources in the early universe.
Decaying primordial black holes, topological defects, the very early
supernovae, super massive particles or structure formation shocks are
some of these \citep{Biermann_introduction,Stanev,Pfrommer}.

In this paper we will investigate whether an early population of
cosmic rays can trigger molecular hydrogen formation and contribute to
the process of small scale structure formation. This will be done by
coupling in a self-consistent way the spectral evolution of a cosmic
ray population with the chemical evolution of a primordial gas. In
Section \ref{prim_chem} and \ref{CR_spec} we describe the chemical 
evolution of the gas and the evolution of the cosmic ray spectrum
respectively. In Section \ref{CR_inter} and \ref{CR_Chem} we describe
how cosmic rays interact with the ambient matter and by which
reactions they influence the primordial chemistry. The method and the
code developed during this work will be described in Section
\ref{Method}. Finally, Section \ref{Discussion} contains a discussion
on the results which will be summarized in Section \ref{Conclusion}.

\section[]{Primordial chemistry}
\label{prim_chem}

As discussed in the Introduction the basic idea of this work is to
study the effect of cosmic rays on a gas consisting of different
primordial chemical species. Among others, we want to investigate the
relevance of cosmic rays for early structure formation, which takes
place in an almost metal free universe between redshift \(z\sim40\)
and \(z\sim10\). We therefore consider only the chemistry of
primordial elements, hydrogen, \(\rm{H}\), helium, \(\rm{He}\) and 
deuterium, \(\rm{D}\). These three species interact via a complicated 
network of chemical reactions, which changes the overall chemical state of the gas. 

According to \citet{Spitzer_book} the velocity distributions of the
gas particles in the density regime considered in this work (\(1\) to
\(10^4\unit{cm^{-3}}\)) are always close to a Maxwellian equilibrium
distribution. This allows to describe the thermal state of this
multi-component gas with just one single kinetic temperature \(T\) for all particles.  

Although the velocity distributions of the particle species are close to that in thermal equilibrium, the level population of the different species does not correspond to thermal equilibrium. In the density regime considered here though, the coronal limit can be applied. This allows to treat all particle species, including molecular hydrogen, in their ground states as long as only moderate UV fields with fluxes up to \(\sim 10^{-15}\unit{erg\,cm^{-2}\,s^{-1}\,Hz^{-1}}\) are present \citep{Abel_chem, Shull_1978}.

Given the above approximations, a simplified approach can be
used to describe numerically the chemical and thermal evolution of the gas.
The chemical evolution of a gas is governed by inelastic collisions that change the number densities of interacting species by creating or destroying chemical compounds. The equation which governs the number density evolution is most generally given as: 
\begin{equation}
\label{chemrateequation:chem_rate_equation}
\frac{dn_i}{dt}=\sum_j\sum_k \alpha ^i k^i_{jk}(T)n_j n_k + \sum_l \beta ^i\lambda^i_l n_l-n_i\frac{d\ln V}{dt}\, .
\end{equation}
Here the subscripts \(i\),\(j\),\(k\) and \(l\) denote the different
particle species, and the quantities \(k^i_{jk}(T)\) and
\(\lambda^i_{l}\) are the rates for the chemical reactions that create
or destroy particles of type \(i\) ( reference to the used rates is
given in Appendix \ref{chemrat}). \(\alpha ^i\) and \(\beta ^i\) are the stoichiometric coefficients that account for the number of particles of type \(i\) that are created or destroyed in each reaction. The last term in equation (\ref{chemrateequation:chem_rate_equation}) describes the change in particle number density due to the evolution of the volume \(V\).

Whereas the temperature evolution of the gas is described by:
\begin{equation}
\label{Tempevolveq:Temp_evolv_eq}
\frac{dT}{dt}= \frac{2(\Gamma-\Lambda)}{3nk_B}-\frac{2}{3}\frac{d\ln V}{dt}T\, ,
\end{equation}
where \(n=\sum_i n_i\) is the total number density and \(\Gamma\) and \(\Lambda\) are the heating and cooling functions respectively (Appendix \ref{coolrat}). The first term on the right hand side of equation (\ref{Tempevolveq:Temp_evolv_eq}) describes the net thermal heat input to the medium, while the second term describes the work done by compression or expansion of the gas. Note that here we assume that the gas behaves as an ideal mono-atomic gas \citep{Spitzer_book}.
Effects of thermal conduction are neglected in equation (\ref{Tempevolveq:Temp_evolv_eq}), which is appropriate for the density and temperature range considered here \citep{Spitzer_book}.

\section{Cosmic ray spectral evolution}
\label{CR_spec}

It is well known that radiation like cosmic rays can ionize neutral
atoms and deeply influence the chemical evolution of a medium.
Since protons are the dominant cosmic ray species at least in our
Galaxy \citep{Pfrommer}, here we will only consider the cosmic ray
protons. This is a fair assumption as nucleosynthesis predicts
\(\approx 93 \% \) of all baryon particles in the universe to be
protons. Therefore the dominant species emitted by some early cosmic
ray accelerator should also be protons. However, a small fraction of
helium might be present in the early cosmic radiation. These \(
\alpha\)-particles can simply be treated as four protons and hence be
absorbed in the proton spectrum \citep{Ensslin_2006}.

 Since the ionization cross section of hydrogen has a maximum at proton
impact energies \(\approx 25.0\unit{keV} \), it is necessary to find
an adequate representation of the cosmic rays at those low
energies. In addition, their original spectrum will be modified due to
the energy losses which cosmic rays experience in the frequent
interactions with particles of the ambient medium. As no measurement
of low energetic cosmic rays is available and the cosmic ray ionization rate depends on the spectrum, we need to model the low energy spectrum and its time evolution theoretically.

In the following we will define some variables useful for our calculations. It is convenient to introduce the dimensionless momentum:
\begin{equation}
\label{dimensionlessmomentum:dimension_lessmomentum}
p=\frac{p_{CR}}{m_pc}\, ,
\end{equation}
where \(p_{CR}\) is the momentum of the cosmic ray proton  and \(m_p\) is the proton mass.
In addition, we can express the kinetic energy \(T_p\) as:
\begin{equation}
\label{kineticenergy:kinetic_energy}
T_p(p)=\left(\sqrt{1+p^2}-1\right)(m_pc^2)
\end{equation}
and the dimensionless velocity \(\beta\) as:
\begin{equation}
\label{dimensionlessvelocity:dimensionless_velocity}
\beta=\frac{v}{c}=\frac{p}{\sqrt{1+p^2}}
\end{equation}
where \(v\) is the particle velocity.

With this notation the spectral evolution equation of the cosmic ray proton number density \(n(p,t)\) can be written as follows:
\begin{eqnarray}
\label{fullpropequation:full_prop:equation}
\frac{\partial n(p,t)}{\partial t} + \frac{\partial \dot{p}(p,t)n(p,t)}{\partial p}=Q(p,t)-\frac{n(p,t)}{\tau(p,t)}\, ,
\end{eqnarray} 
where the dot denotes the derivative with respect to time, \(Q(p,t)\) is a source
function for the cosmic ray protons and \(\tau(p,t)\) is a typical
loss time for cosmic ray particles from the volume under
consideration.

Under the assumption that the injected spectrum obeys a power law with some spectral index \(s\) in momentum space, the source function can be written as:
\begin{equation}
\label{sourceterm:source_term}
Q(p,t)=\frac{\dot{\epsilon}_{CR}}{m_pc^2\,\kappa(s,p_0)}p^{-s}\Theta(p-p_0)\, ,
\end{equation}
where \(\dot{\epsilon}_{CR}\) is the change of total cosmic ray energy density with respect to time, \(\kappa(s,p_0)\) is a normalization depending on the spectral index \(s\), \(\Theta(x)\) is the Heaviside function and \(p_0\) is a lower momentum cutoff of the injected spectrum. The value of \(\kappa(s,p_0)\) can be calculated as \citep{Ensslin_2006}:
\begin{equation}
\label{kappas:kappa_s}
\kappa(s,p_0)=\frac{1}{s-1}\left(p_0^{1-s}\left(\sqrt{1+p_0^2}-1\right)+\frac{1}{2}B_{\frac{1}{1+p_0^2}}\left(\frac{s-2}{2},\frac{3-s}{2}\right)\right)\,
\end{equation}
where \(B_x(a,b)\) denotes the incomplete Beta-function, and \(s>2\) is assumed.

\section{Interaction between cosmic rays and matter}
\label{CR_inter}

\subsection {Coulomb losses}

Here we are interested in the problem of the energy loss of cosmic
rays in the interstellar and intergalactic gas. The basis of all
electromagnetic interactions is the Coulomb scattering between
electric charges, which allows for transfer of kinetic energy as the proton moves through a free electron gas. According to \citet{Gould}, the total energy loss of a proton by Coulomb losses in a plasma is given by:
\begin{equation}
\label{coulombloss:coulomb_loss}
-\left(\frac{dT_{p}(p)}{dt} \right)_{C}=\frac{4\pi e^4n_{e}}{m_{e}\beta c}\left[\ln\left(\frac{2m_ec^2\beta p}{\hbar \omega_{pl}}\right)-\frac{\beta^2}{2} \right]\, .
\end{equation}
Here, \(\omega _{pl}=\sqrt{4\pi e^2n_e/m_e}\) is the plasma frequency
and \(n_{e}\) is the number density of free electrons. The result is
independent of the mass \(m_{p}\) of the incident proton as long as
the recoil of the proton can be neglected, i.e. \(\gamma m_{e}
c\ll m_{p}c\), with \(\gamma = 1/ \sqrt{1-\beta^2}\). Thus equation (\ref{coulombloss:coulomb_loss}) is valid for \(\gamma\ll m_{p}/m_{e}\). In a partially ionized medium the total loss of kinetic energy can simply be calculated by adding the losses due to excitation or ionization to the Coulomb losses as calculated here \citep{Gould}.

\subsection {Ionisation losses}

Ionization energy losses are important for all charged particles, and for particles other than electrons and positrons they dominate over radiation energy losses at all but the highest energies. These ionization losses are governed by the Bethe-Bloch equation \citep{Groom_Klein}, which for moderately relativistic charged particles assumes the form:
\begin{eqnarray}
\label{ionloss:ion_loss}
-\left(\frac{dT_{p}(p)}{dt} \right)_{I}&=&\frac{4\pi e^4}{m_{e}\beta c}\sum_{Z}Zn_{z}L_{0} \nonumber \\
                                       &=&\frac{4\pi e^4}{m_{e}\beta c}\sum_{Z}Zn_{z}\left[\ln \left(\frac{2m_ec^2p^2}{I_{Z}b(\gamma)}\right)-\beta ^2-\frac{\delta _{Z}}{2}\right]
\end{eqnarray}
where \(n_{Z}\) is the number density of atomic species with electron number \(Z\), \(I_{Z}\) is the ionization potential, and \(L_{0}\) is called the stopping number \citep{Ziegler_1999}. The quantity \(b(\gamma)=\sqrt{1+2\gamma m_e/m_p + (m_e/m_p)^2}\) is a correction factor that accounts for the maximum kinetic energy that can be imparted to a free electron in a single collision \citep{Groom_Klein}.

The density correction factor \(\delta _{Z}\) accounts for the
screening effect of a medium which becomes polarized as highly
relativistic particles move through it \citep{Martin_Shaw}. In this 
case the electromagnetic field of the ions may not be at the assumed 
free-space value, but is reduced by the dielectric constant of the
medium \citep{Ziegler_1999}. The density correction factor \(\delta
_{Z}\) is usually negligible for gases \citep{Groom_Klein}, but is given here for completeness:
\begin{eqnarray}
\label{denseffect:dens_effect}
\delta _{Z}&=& \left\{\begin{array}
								{r@{\quad:\quad}l} 
								2y-D_Z & (y_{1,Z}<y) \\ 
								2y-D_Z+a_Z((y_{1,Z}-y)/ln10)^{k_{Z}}& (y_{0,Z}<y<y_{1,Z}) \\ 																		0 & (y<y_{0,Z}) \\
										\end{array}\right. 
\end{eqnarray}
Here, \(y=\ln(p)\), \(D_Z=1-2\ln(\hbar \omega _{pl}/I_Z)\) and \(y_{0,Z}\), \(y_{1,Z}\), \(a_Z\), \(k_Z\) are empirical constants which characterize the atomic species \citep{Sternheimer_1952} and are given in table \ref{tab:Ionizationpotential} together with the values for the ionization potentials \(I_{Z}\) of molecular hydrogen and helium. All hydrogen measurements were done with molecular hydrogen, but not much difference is expected when these values are adopted for atomic hydrogen.

As already pointed out earlier, the energy range of cosmic ray protons
most important for the primordial chemistry is of order
\(\unit{keV}\). The Bethe-Bloch equation in the form
(\ref{ionloss:ion_loss}) is not valid at these low energies, and a
correction term must be added to the square brackets of equation (\ref{ionloss:ion_loss}). The so called Bloch correction is important for slow particles \citep{Mannheim_Schlickeiser}, and for protons it can be written as:
\begin{equation}
\label{Blochcorrection:Bloch_Correction}
L_{Bloch}=\frac{1}{2}\left(\Psi(1)-\rm{Re}\Psi\left(1+\frac{i\alpha}{\beta}\right)\right)
\end{equation}
where \(\Psi(x)\) is the digamma function, the logarithmic derivative of the gamma function, and \(\alpha=1/137\) is the fine structure constant \citep{Ziegler_1999}.

With this corrections the Bethe-Bloch equation can be used for particles with velocities greater than the characteristic velocity of the medium's electrons, which is the orbital velocity in the K-shell of the atom  \citep{Mannheim_Schlickeiser}. For atomic hydrogen the orbital velocity can be derived from the Bohr atomic model and yields:
\begin{equation}
\label{orbitalvel:orbital_vel}
\beta _{orbit}= \frac{e^2}{\hbar c}\approx 0.0073\, .
\end{equation}
For protons with velocities larger than \(\beta _{orbit}\) the Bethe-Bloch equation is in excellent agreement with the experimental values \citep{Mannheim_Schlickeiser}.

At even lower particle velocities Ginzburg and Syrovatskii provide a useful ionization loss rate for slow ions \citep{Ginzburg}, which for protons can be written as:
\begin{equation}
\label{ionloss_Ginzburg:ion_loss_Ginzburg}
-\left(\frac{dT_{p}(p)}{dt} \right)_{I}=2.11 \times 10^{-2} \beta ^2 \sum _Z n_Z Z \unit{\frac{eV}{s}}\, .
\end{equation}
For proton velocities lower than \(\beta _{orbit}\) this agrees well with the experimental data of \citet{Whaling}. Then the total ionization energy loss formula is given by:
\begin{eqnarray}
\label{ion_energyloss:ion_energy_loss}
-\left(\frac{dT_{p}(p)}{dt} \right)_{I}&=& \left\{\begin{array}
								                      {r@{\quad:\quad}l} 
								                       \frac{4\pi e^4}{m_{e}\beta c}\sum_{Z}Zn_{z}\left[L_{0}+L_{Bloch}\right] & (\beta>\beta _{orbit}) \\ 
								                      2.11 \times 10^{-2} \beta ^2 \sum _Z n_Z Z \unit{\frac{eV}{s}} & (\beta<\beta _{orbit})\, .\\                                                         
										\end{array}\right. 
\end{eqnarray}
\begin{table}
	\centerline{
	\begin{tabular}{ccclllll}
	\hline
	Element & Z & \(I_{Z}\) \(\left [\unit{eV}\right ]\) & \(y_{0,Z}\) & \(y_{1,Z}\) & \(a_Z\) & \(k_Z\) \\
	\hline
	\(\rm He\)&2& \(24.6 \) & \(2.0\ln10\) & \(3.0\ln10\) & \(0.98\) & \(4.11\)\\
	\(\rm H_{2}\)&1 & \(13.6\)& \(1.76\ln10\) & \(3.0\ln10\) & \(0.34\) & \(5.01\)\\
	\hline
	\end{tabular}}
	\caption{Atomic data of \(\rm He\) and \(\rm H_{2}\) as given by \citet{Sternheimer_1952}. The hydrogen measurements were done with molecular hydrogen, but no significant changes for atomic hydrogen are expected within the accuracy needed for our purpose \citep{Ensslin_2006}.}
	\label{tab:Ionizationpotential}
	\end{table}

Note that the derivation of the cosmic ray energy losses due to
ionization neglects the details of the atomic shell structure. In
order to obtain accurate energy losses, the particle electron
collisions should be considered with detailed treatment of each target
electron's orbital bonding, and as the particle slows down, a shell
correction term should be added to equation (\ref{ionloss:ion_loss}) 
\citep{Ziegler_1999}. All the atoms considered in this work though, on average will always be in the ground state and thus no detailed study of the atomic shell corrections is needed. The correction would be at most \(6\%\) in the energy range \(1-100\unit{MeV}\) \citep{Ziegler_1999}. Therefore, shell correction can be safely neglected, and equation (\ref{ion_energyloss:ion_energy_loss}) describes adequately the ionization losses of cosmic ray particles.

\subsection {Hadronic losses}

The known cosmic ray spectrum extends over energies from a few hundred\(\unit{MeV}\) to \(3\times 10^{20}\unit{eV}\) \citep{Biermann_introduction}. Such high relativistic energies allow cosmic ray protons to overcome the Coulomb barrier and interact hadronically with the nuclei of the ambient matter. The fundamental theory underlying high energy collisions of hadrons is generally referred to as Quantum Chromo Dynamics (QCD) \citep{Stanev}. As QCD predicts the creation of massive particles in collisions between hadrons if the incident particle possesses enough kinetic energy \citep{peskin}, these interactions, especially pion creation, are an important energy loss mechanism for cosmic ray protons at high energies. The lightest known mesons are pions, with masses \(m_{\pi^{\pm}}c^2=139.57\unit{MeV}\) for charged pions and \(m_{\pi^{0}}c^2=134.98\unit{MeV}\) for neutral pions \citep{Martin_Shaw}. For positive and neutral pion production the lightest final states which conserve baryon number and electric charge are created by the processes
\begin{equation}
\label{pospionprod:pos_pionprod}
\rm{P}+\rm{P}\to \rm{P}+\rm{N} + \pi^+
\end{equation}  
and
\begin{equation}
\label{neutpionprod:neut_pionprod}
\rm{P}+\rm{P}\to \rm{P}+\rm{P} + \pi^0
\end{equation}  
yielding a threshold energy \(E_{CR}^0\approx1.22\unit{GeV}\), with a corresponding momentum threshold \(p_{th}m_pc^2\approx0.79\unit{GeV}\) \citep{Martin_Shaw}.

The neutron produced in reaction (\ref{pospionprod:pos_pionprod}) will decay into a proton after a mean lifetime of \(886\unit{s}\), so we basically end up with pions and two protons.
This assures that the cosmic ray proton number density is conserved in these reactions. In subsequent processes the pions will decay via the channels 
\begin{equation}
\label{chargedpiondecay:charged_piondecay}
\pi^{\pm}\to\mu^{\pm} +\nu _{\mu} / \bar{\nu _{\mu}} \to e^{\pm} +\nu _{e}/\bar{\nu _{e}}+\nu _{\mu}/\bar{\nu _{\mu}}
\end{equation} 
and
\begin{equation}
\label{neutralpiondecay:neutral_piondecay}
\pi^{0}\to 2\gamma
\end{equation}
with mean lifetimes for charged pions of \(2.6\times 10^{-8}\unit{s}\) and for neutral pions of \(8.4\times 10^{-17}\unit{s}\) \citep{Martin_Shaw}. However, the subsequent decay products are not considered further in this work.

The average kinetic energy loss of cosmic ray protons due to pion production has been calculated by   \citet{Mannheim_Schlickeiser} and can be written in the form
\begin{equation}
\label{hadronicloss:hadronic_loss}
-\left(\frac{dT_{p}(p)}{dt} \right)_{had}=0.65 c n_{nuclei} \sigma_{pp}T_{p}\Theta(p-p_{th})\, ,
\end{equation}
where \(n_{nuclei}\) is the number density of the target nucleon in the medium and \(\sigma_{pp}=3\times10^{-26}\unit{cm^2}\) is the cross section for pion production in proton collisions.

\section{Coupling between cosmic ray protons and primordial gas chemistry}
\label{CR_Chem}

In this Section we describe the method used and the approximations
adopted to couple cosmic ray protons to the chemical evolution of a primordial gas.

\subsection{Ionization rates}
\label{subsec:ionrate}

As a cosmic ray proton travels through matter it leaves behind a trail of ions. In accelerator experiments it is possible to measure the mean energy needed to create one electron-ion pair \(W_Z\) along the track, which clearly depends on the target material \citep{Bakker_1951}. According to \citet{Heitler} this average energy loss is representative for the average primary ionization of the cosmic ray proton.

Both the fractional number of cases in which a collision with an atom results in ionization rather than excitation to a discrete level and the average energy transferred to the ionized electron, are almost independent of the incident energy of the cosmic ray proton \citep{Mannheim_Schlickeiser}. With these empirical observations and equation (\ref{ion_energyloss:ion_energy_loss}), the primary ionization rate for the chemical elements can be calculated as:
\begin{equation}
\label{ionizationrate:ionization_rate}
\lambda_{CR}^{prim}=\frac{\dot{\epsilon}_{I}(p_{min})}{W_Z\,n_Z} \, ,
\end{equation}
where \(\dot{\epsilon}_{I}(p_{min})\) is the change of ionization energy density per time given as:
\begin{equation}
\label{ionizationratepervolume:ionization_rate_pervolume}
\dot{\epsilon}_{I}(p_{min})=\int_{p_{min}}^{\infty} \left|\left(\frac{dT_{p}(p)}{dt} \right)_{I} \right | n(p) dp								\, . 
\end{equation}
Here \(p_{min}\) is defined by the minimum energy necessary to ionize the atomic species (in the case of atomic hydrogen it is \(13.6\unit{eV}\)) and \(W_Z\) is the mean energy expended per ion pair created by a proton. The values of \(W_Z\) were measured by \citet{Bakker_1951} and  \citet{Weiss_Bernstein} for several elements, and are given in table \ref{tab:WZ} for molecular hydrogen and helium. The hydrogen measurements were done with molecular hydrogen, but no significant changes for atomic hydrogen are expected within the accuracy required here.

When a cosmic ray proton ionizes an atom, energy is transferred to the ejected electron as kinetic energy. Some of this energy will go into further ionization or excitation of atoms, followed by radiation, while some of the energy will go directly into heating the medium \citep{Spitzer_Scott}. Therefore, this first generation electrons give rise to additional ionization and more free electrons as long as their kinetic energy is above the ionization threshold energy.

According to \citet{Spitzer_Scott} all secondary and higher generation
ionization can be accounted for multiplying the primary ionization rate (\ref{ionizationrate:ionization_rate}) by a factor \(\xi=5/3\). This holds up to an ionization fraction of \(f=n_e/n_H\approx10^{-2}\), while for higher ionization fractions secondary electrons loose their energy to Coulomb interactions with the free electron gas rather than to ionization of additional atoms \citep{Spitzer_Scott}. As cosmic rays can raise the ionization fraction considerably, especially for low density gases, one should be aware of the errors that are introduced by using a constant \(\xi\). However, at this stage of research we are mainly interested in finding an overall effect and therefore we treat \(\xi\) as a parameter. Comparison of calculations with \(\xi=5/3\) and \(\xi=1\)  showed that not much difference arises from the different treatment of secondary electrons in the cases studied here \citep{Jens_Diplom}.

With all the above simplifications the total ionization rate, including primary and higher generation ionizations, can be written as:
\begin{equation}
\label{totalionization:total_ionization}
\lambda_{CR}^{tot}=\xi\lambda_{CR}^{prim}\, .
\end{equation}

\subsection{Cosmic ray input on the chemical network}

The most important reactions by which cosmic ray protons influence the primordial chemical network are the following:
\begin{enumerate}
\item \(\rm{P_{CR}+H\to P_{CR}+H^{+}+e^{-}}\) 
\item \(\rm{P_{CR}+He\to P_{CR}+He^{+}+e^{-}}\)
\item \(\rm{P_{CR}+D\to P_{CR}+D^{+}+e^{-}}\)
\item \(\rm{P_{CR}+H_{2}\to P_{CR}+H_{2}^{+}+e^{-}}\) 
\item \(\rm{P_{CR}+H^{-}\to P_{CR}+H+e^{-}}\)
\end{enumerate}
where the electron detachment reaction (v) turned out to be negligible \citep{Jens_Diplom}.

All these reactions conserve the cosmic ray proton number density, and therefore no exchange of particles between cosmic rays and medium needs to be considered. The ionization rates for the reactions can be calculated with equation (\ref{totalionization:total_ionization}) and the data taken from tables \ref{tab:Ionizationpotential} and \ref{tab:WZ}. Unfortunately, no experimental data for deuterium targets is available, and therefore we use the same values as for hydrogen, which should be a safe assumption for the accuracy required in this work. As \(97\%\) of cosmic ray-\(\rm{H_2}\) impacts lead to the formation of \(\rm{H_{2}^+}\) via reaction (iv) no other destruction mechanism for molecular hydrogen by cosmic ray protons will be considered here in agreement with the literature \citep{Hartquist_1995}. However, inclusion of additional cosmic ray-\(\rm{H_2}\) reactions lead to the formation of additional \(\rm{H_2}\). Such reactions and also the influence of cosmic rays on other molecules like \(\rm{HD}\) will be discussed in detail in forthcoming publications.

 Whenever there is helium present in the medium, it will not only be ionized by the cosmic ray protons, but will also emit a \(21.23\unit{eV}\) photon, as the remaining electron relaxes to a new shell configuration \citep{Spitzer_Scott}. In principle this photon is capable of ionizing further atomic hydrogen. But since we consider a very rarefied medium and the flux of these photons is assumed to be low, they are expected to contribute insignificantly to the overall ionization rate.
\begin{table}
	\centerline{
	\begin{tabular}{llll}
	\hline
	Element & Z & \(W_{Z}\) \(\left[\unit{eV}\right]\) \\
	\hline
	\(\rm He\)&2& \(40.3 \pm 0.8\)\\
	\(\rm H_{2}\)&1 & \(36.3\pm 0.7\)\\
	\hline
	\end{tabular}}
	\caption{Experimental values of \(W_{Z}\) for molecular hydrogen \citep{Bakker_1951} and helium \citep{Weiss_Bernstein}.}
	\label{tab:WZ}
	\end{table}

\subsection{Heating rates}
\label{CRheating:CR_heating}

Cosmic rays can be an efficient heat source especially for a low density gas, and therefore their heat input to the medium must be considered. When the cosmic ray proton ionizes an atom, it transfers a certain amount of kinetic energy to the electron, which is either used for further excitation and ionization of atoms, or is distributed by elastic collisions to other species of the medium. In the latter case the overall kinetic temperature is raised.

The effect of heating by cosmic ray ionization and secondary electrons is not well described in the literature. The papers that mention heating by cosmic ray ionization are not very detailed and show some discrepancies (e.g. \citet{Mannheim_Schlickeiser,Spitzer_Scott}). Here we will provide a reasonable upper limit to the heat input.

Let \(E_{ion}^Z\) be the ionization threshold energy of an atom of atomic charge \(Z\) (for hydrogen \(E_{ion}^1=13.6\unit{eV}\)). Then the mean kinetic energy \(E_{kin}\) imparted to the primary electron in a single collision can be calculated as
\begin{equation}
\label{meanelenergy:mean_el_energy_per_sec}
E_{kin}=\left(W_Z-E_{ion}^Z\right)\, .
\end{equation}
In subsequent collisions the primary electron will on average ionize \(\xi-1\) atoms and therefore loose the additional kinetic energy \((\xi-1)E_{ion}^Z\). Thus every primary ionization by a cosmic ray particle finally yields a temperature gain equivalent to an energy \(E_{heat}=\left(W_Z-\xi E_{ion}^Z\right)\). Hence we can write the heating rate for cosmic ray ionization as 
\begin{equation}
\label{CRionheatingrate:CR_ion_Heating_rate}
\Gamma_{CR}^{ion}=\left(W_Z-\xi E_{ion}^Z\right)\lambda_{CR}^{prim}n_{Z}=\left(\frac{W_Z}{\xi}- E_{ion}^Z\right)\lambda_{CR}^{tot}n_{Z}
\end{equation}
which, for \(\xi=5/3\), means a heat input of \(E_{heat}=\approx 8.18 \unit{eV}\) for every ionization of hydrogen in a neutral medium. 
This heat input is an overestimation as the electron could have also lost energy to excitation which was not accounted for here.
Thus, in a neutral medium less than \(1/4\) of the energy transferred from the cosmic ray proton is finally deposited as heat. Again, it should be noted that \(\xi\) decreases with increasing ionization fraction \(f\) and thus the heat input by cosmic rays will increase with \(f\).
As soon as a large fraction of the medium is ionized, the cosmic ray proton looses kinetic energy mainly due to Coulomb interactions with the free electrons of the medium, and all the kinetic energy imparted to the free electrons is equivalent to the heating energy \(E_{heat}\).

The heating rate due to Coulomb interactions can be calculated by using equation (\ref{coulombloss:coulomb_loss}):
\begin{equation}
\label{Coulombheat:Coulomb_Heating_rate}
\Gamma_{CR}^{coul}=\int_{0}^{\infty}\left|\left(\frac{dT_{p}(p)}{dt} \right)_{C}\right|n(p)\,dp\, .
\end{equation}
The fact that the energy lost by cosmic ray protons in Coulomb interactions is directly transferred to heat of the medium is the reason why cosmic rays are an efficient heat source. Thus, once the gas is ionized to sufficiently high degrees the future fate of the medium will be defined by Coulomb heating solely. 

\section{Method}
\label{Method}

Here we will describe and discuss the numerical methods used to study the influence of early cosmic rays on primordial gas. 
In order to estimate the effects of cosmic ray protons on the chemistry of primordial gas, it is necessary to follow the chemical and thermal evolution of such gas along with the cosmic ray spectral evolution. Therefore we simulate a homogeneous and isotropic medium and follow the evolution of 13 species (\(\rm{H}\), \(\rm{H^+}\), \(\rm{H^-}\), \(\rm{H_2^+}\), \(\rm{H_2}\), \(\rm{He}\), \(\rm{He^+}\), \(\rm{He^{++}}\), \(\rm{HeH^+}\), \(\rm{D}\), \(\rm{D^+}\), \(\rm{HD}\) and \(\rm{e^-}\)) self-consistently together with the spectral evolution of the cosmic ray protons. The code calculates the chemical evolution of primordial gas by taking into account 27 collisional rates ( Appendix \ref{chemrat} ), 17 radiative rates including CMB interactions ( Appendix \ref{chemrat} ) and 4 additional rates arising from the coupling of cosmic ray protons to the chemical network as mentioned earlier.

\subsection{Code testing}

The accuracy of the numerical scheme for the cosmic ray spectral evolution is checked by comparing the numerical results for the cosmic ray spectral evolution with the analytic equilibrium solution of equation (\ref{fullpropequation:full_prop:equation}) where we neglect the escape term.
The deviation of the numerical results from the analytic ones is measured by
\begin{equation}
\label{comparaanalytic:compare_analytic}
\delta n(p)= \frac{n_{analytic}(p)-n_{numerical}(p)}{n_{analytic}(p)} \mbox{.}
\end{equation}
Here we do not consider any chemical evolution and assume a gas consisting of atomic hydrogen with a number density \(n_{\rm{H}}=10^2\unit{cm^{-3}}\). The cosmic ray spectrum is integrated from the lower boundary \(p_{min}=10^{-3}\) to the upper boundary \(p_{max}=10^{3}\) with a spectral index \(s=2.2\) and a cosmic ray energy density injection rate \(\dot{\epsilon}_{CR}=10^{-26}\unit{erg\,cm^{-3}\,sec^{-1}}\). 
\begin{figure*}
	\centering
		\rotatebox{270}
	{
		\includegraphics[bb = 80 120 590 740,width=0.7\textwidth,clip=true]{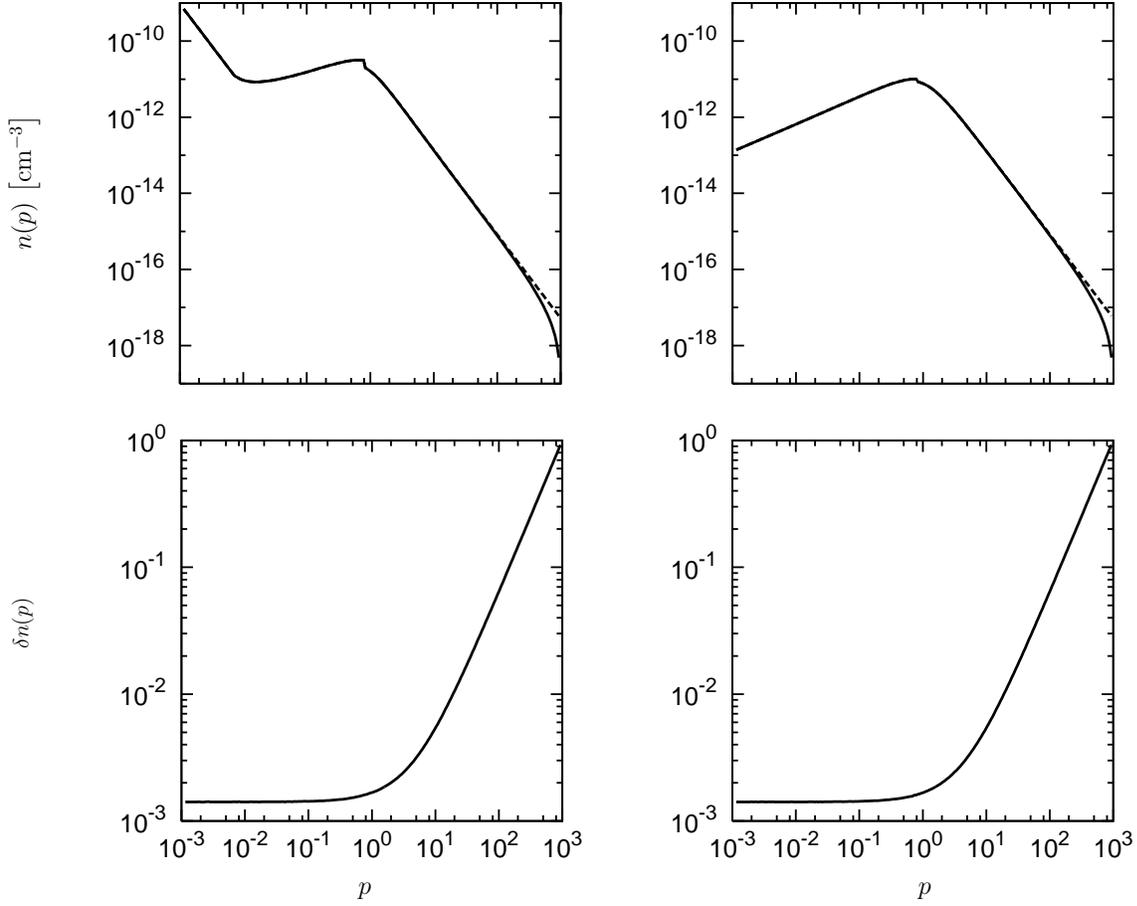}
	}
	\caption{Upper panels: cosmic ray spectrum after interaction with a neutral (left panel) and an ionized (right panel) gas. The dashed (solid) lines indicate the analytic (numerical) results. Lower panels: \(\delta n(p)\) for a neutral (left panel) and an ionized (right panel) gas.}
	\label{fig:codetestCR}
\end{figure*}

The asymptotic steady state solutions are then compared for two scenarios. In the first scenario we consider that the cosmic rays interact with a totally neutral medium, while in the second we choose a totally ionized gas. The results are plotted in figure \ref{fig:codetestCR}. The numerical solutions are nearly identical to the analytic ones. At the highest momenta the boundary effects lead to some deviations, which are still within the accuracy needed for this work. In any case these will not affect the calculations considered here since those high momenta are far beyond the region of interest for cosmic ray ionization. Therefore the calculations done in this work will not suffer from accuracy losses due to numerical effects.

Testing the chemical network is more difficult as no analytic solution exists. For this reason, we have chosen to check the behavior of the code against results present in the literature. We run a simulation to follow the chemical evolution of the intergalactic medium from redshift \(10^4\) to \(0\) and compare the results with those of \cite{Galli_Palla}.

\begin{figure*}
	\centering
		\rotatebox{270}
	{
		\includegraphics[bb = 70 140 910 700,width=1\textwidth,clip=true]{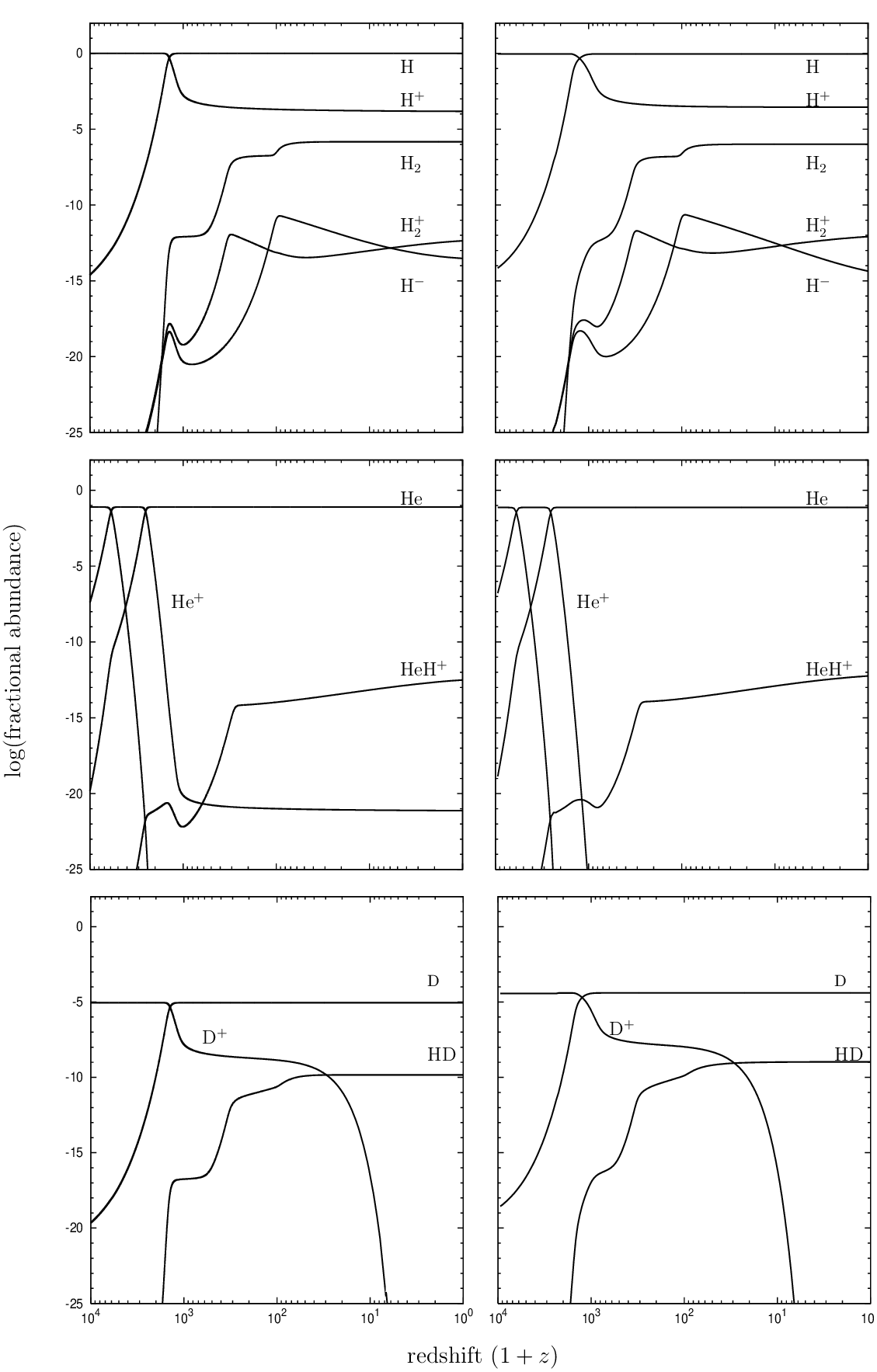}
	}
	\caption{Chemical evolution of a primordial gas described in this work (left panels) and in \citet{Galli_Palla} (right panels).}
	\label{fig:chem_test}
\end{figure*}

For our calculations we adopt a \(\Lambda \rm{CDM}\) cosmological model with \(\Omega _{\Lambda}=0.73\), \(\Omega _{m}=0.27\) and \(H_{0}=71 \unit{km\,sec^{-1}\, Mpc^{-1}}\). As initial conditions we assume a completely ionized gas with mass fractions \(f_{\rm{H}}=0.76\), \(f_{\rm{He}}=0.24\) and \(f_{\rm{D}}=1.8\times 10^{-5}\,f_{\rm{H}}\), as predicted by nucleosynthesis \citep{smith_initial}. The result is presented in figure \ref{fig:chem_test}.
From a comparison between  the results obtained in this work and those of Galli and Palla, it is clear that the main features of the chemical evolution are very well reproduced by the code, although some minor differences are present. These arise because of the slightly different sets of chemical rates used and because Galli and Palla included more chemical species.

Finally as chemical reactions conserve the mass of the system we checked that the mass remains constant throughout the calculations.

\subsection{The simulation}

The gas in the simulations consists of the primordial elements \(\rm{H}\), \(\rm{D}\) and \(\rm{He}\) with the mass fractions \(f_{\rm{H}}=0.76\), \(f_{\rm{He}}=0.24\) and \(f_{\rm{D}}=1.8\times 10^{-5}\,f_{\rm{H}}\) \citep{smith_initial}.

The parameters that determine the cosmic ray spectral evolution are set as follows. We set the boundaries for the integration region in momentum space at \(p_{min}=10^{-3}\) and \(p_{max}=10^{4}\), and assume the momentum cutoff \(p_0\) in equation (\ref{sourceterm:source_term}) to coincide with \(p_{min}\). The spectral index \(s\) of the injected power-law spectrum is assumed to be \(s=2.2\), as expected for Fermi-acceleration at shocks \citep{Mannheim_Schlickeiser}. This gives \(\kappa(s,p_{min})=4.95\).
The upper boundary \(p_{max}\) is much higher than the range of momentum expected to be important for the chemical network, and therefore boundary effects arising from the numerical integration of the cosmic ray spectrum will not interfere with the calculations of the chemical network. In order to restrict the number of parameters, the escape of cosmic ray particles from the medium under consideration is assumed to be negligible and therefore their escape time is set to \(\tau=\infty\). This implies that either the gas occupies an infinite volume, or magnetic fields capable of confining the cosmic rays in the region of interest are present. As already discussed in Section \ref{subsec:ionrate} in the following we will usually adopt \(\xi=5/3\) and a mean heat input per ionization of \(E_{heat}=8.18\unit{eV}\) for hydrogen and \(4.0\unit{eV}\) for helium. 

As not much about primordial cosmic ray sources and the regions in which cosmic rays might be present in the early universe is known, the code was used to run a large number of simulations in order to explore a wide range of the \(\dot{\epsilon}_{CR}\)\(\, \times \, \)\(n\) parameter space.

Tests with an additional UV photon background field, as it might arise from the very first stars, showed that in the presence of ionizing photons the influence of cosmic rays is usually sub-dominant. We therefore consider only regions that are shielded from ionizing photon radiation, but the treatment of \(\rm{H_2}\) dissociating photons with energies in the range of the Lyman and Werner transitions (\(11.2\)-\(13.6\unit{eV}\)) is included \citep{book_interstellar_medium}. As only this small range of photon energy is of interest, we may assume a constant photon flux \(F\). In the following we will refer to these dissociating photons as soft ultraviolet (SUV) radiation. No radiative transfer of photons is considered in this work, and therefore the dissociating background flux must be understood as a mean flux of photons in the gas. In reality, the molecules in the outer regions of the gas cloud absorb some of the dissociating photons resulting in a decrease of the dissociating radiation toward the inner regions \citep{book_interstellar_medium}. This self-shielding mechanism is clearly dependent on the gas density and distribution and therefore the mean flux of photons in the medium will in general depend on the total number density \(n\) and the sources that produce it.

\begin{figure*}
	\centering
		\rotatebox{270}
	{
		\includegraphics[bb = 80 40 590 670,width=0.7\textwidth,clip=true]{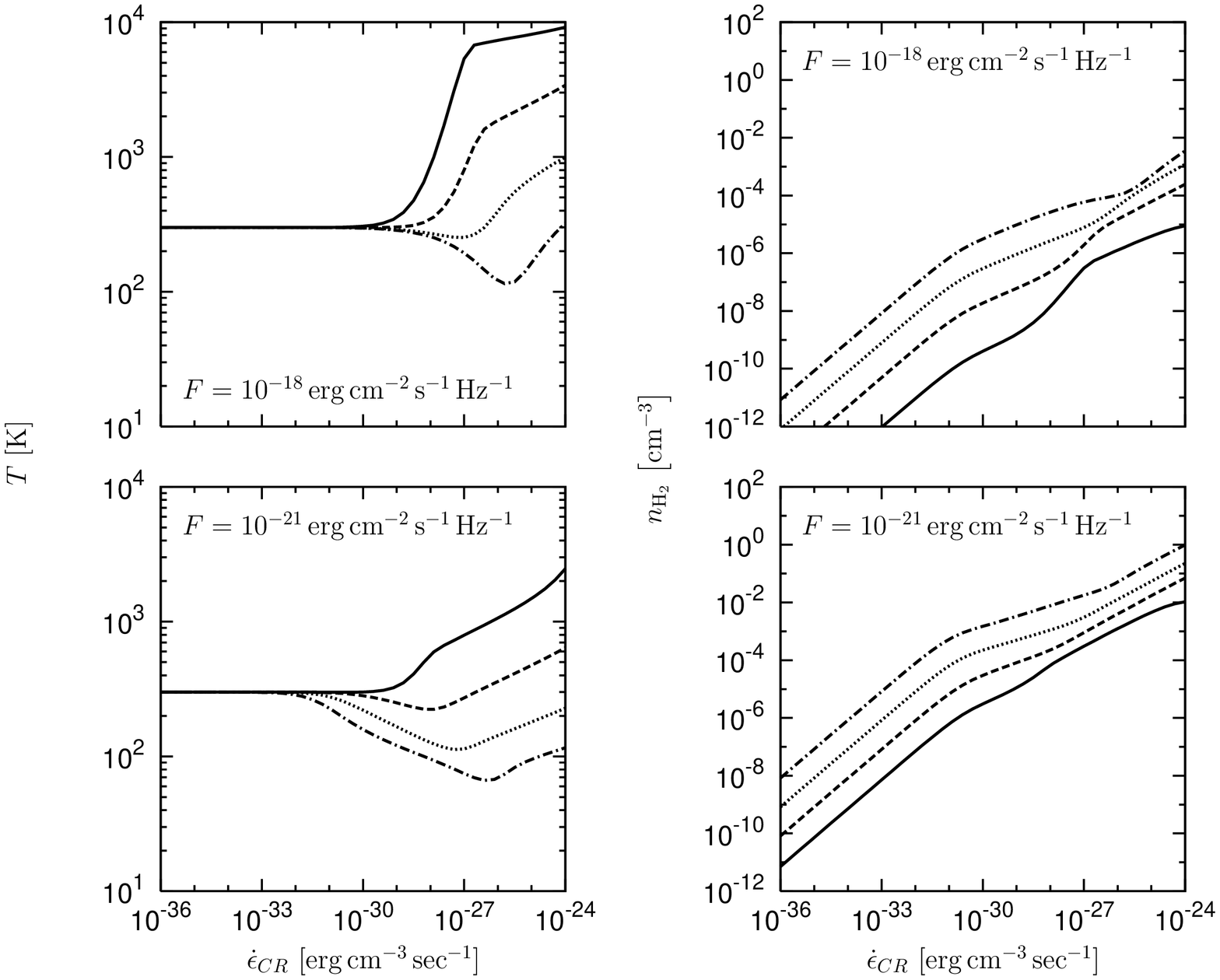}
	}
	\caption{The left panel shows the final temperature and the right panel displays the final \(\rm{H_2}\) number density as a function of \(\dot{\epsilon}_{CR}\) for two different SUV background fluxes. Different curves refer to different total number densities: \(n=1\) (solid line), \(10\) (dashed line), \(100\) (dotted line) and \(1000\unit{cm^{-3}}\) (dash-dotted line). The initial temperature is \(T_0=300\unit{K}\).}
	\label{fig:Temp_300K}
\end{figure*}

\begin{figure*}
	\centering
		\rotatebox{270}
	{
		\includegraphics[bb = 80 40 590 670,width=0.7\textwidth,clip=true]{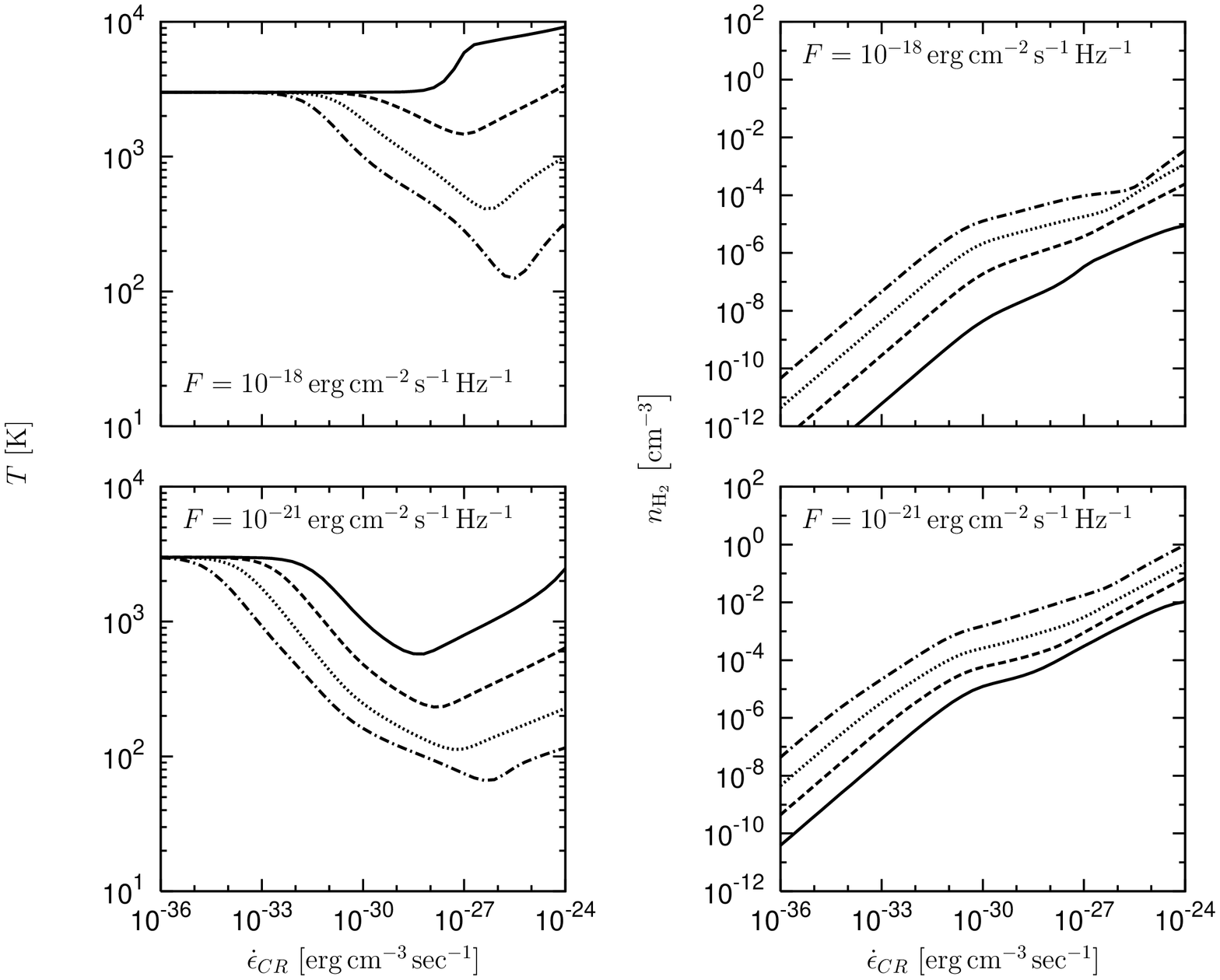}
	}
	\caption{Same as figure \ref{fig:Temp_300K} but for initial temperature \(T_0=3000 \unit{K}\).}
	\label{fig:Temp_3000K}
\end{figure*}

\section{Results}

In this Section we will discuss the effects of CR protons on the chemistry of a primordial gas as derived by our simulations.

It should be noted that cosmic rays are not able to destroy an initially available population of molecular hydrogen. As the electron detachment from \(\rm{H^-}\) by cosmic rays is negligible the only possible destruction mechanism for molecular hydrogen is \(\rm{H_2}\) ionization. Nevertheless, as soon as cosmic rays are strong enough to sufficiently ionize molecular hydrogen, they also provide enough free electrons by ionizing the atomic species, resulting in a net production of H$_2$ \citep{Jens_Diplom}. Therefore, cosmic rays do not interfere with any other chemical process that leads to the formation of molecular hydrogen. This suggests that the final abundance of H$_2$ can be written as:
\begin{equation}
\label{equ:tot_nH2}
n_{\rm{H_2}}\approx n^0_{\rm{H_2}}+\Delta n_{\rm{H_2}} 
\end{equation}
where \(n^0_{\rm{H_2}}\) is the initial number density of molecular hydrogen, and 
\(\Delta n_{\rm{H_2}}\) is the number density produced by cosmic rays in a purely atomic gas of the same total number density.

The simple dependence of the overall behavior of \(n_{\rm{H_2}}\) on its initial abundance allows to make some more general considerations. Let us assume to have a gas cloud with initial \(\rm{H_2}\) abundance \(n^0_{\rm{H_2}}\ge0\) in thermal and chemical equilibrium in the absence of cosmic rays. If \(\rm{H_2}\) line cooling is the main cooling mechanism, this implies that \(\Gamma=\Lambda_{\rm{H_2}}(n^0_{\rm{H_2}})\), where \(\Gamma\) includes all heating sources with exception of cosmic rays. When cosmic rays interact with the gas cloud, its temperature change can be written as:
\begin{eqnarray}
\label{equ:change_temp}
\frac{3}{2}\,n\,k_B\, \Delta T & = & \Gamma+\Gamma_{CR}-\Lambda _{\rm{H_2}} \nonumber \\
& = & \Gamma+\Gamma_{CR}-\left [\Lambda_{\rm{H_2}}(n^0_{\rm{H_2}})+\Lambda_{\rm{H_2}}(\Delta n_{\rm{H_2}}) \right] \nonumber \\
& = & \Gamma_{CR}-\Lambda_{\rm{H_2}}(\Delta n_{\rm{H_2}})\, ,
\end{eqnarray}
if we assume that all additional heating sources remain constant and equal to the initial \(\rm{H_2}\)-cooling. Here we have used the fact that \(\Lambda_{\rm{H_2}}\) in thermal and chemical equilibrium is proportional to the \(\rm{H_2}\) number density.
This implies that the temperature change induced by the interaction of cosmic rays and a gas with any initial \(\rm{H_2}\) abundance can be estimated simply from the balance between \(\Gamma _{CR}\) and the \(\rm{H_2}\) cooling of an initially atomic gas. The advantage is that we can explore a wider physical condition range without knowing the exact initial abundance of \(\rm{H_2}\) and the details of heating sources other than cosmic rays. 

Note that this separation of the chemical network into a purely atomic part, which is influenced by the cosmic rays, and an unchanged part with initial molecular hydrogen, must not be understood as a rule. This approximation is only valid under the assumptions presented here, and if the number density of atomic hydrogen does not change much.

Therefore, we will start our calculations from a neutral purely atomic gas of primordial composition, and all molecular hydrogen \(\rm{H_2}\) will be understood as the additionally produced molecular hydrogen \(\Delta n_{\rm{H_2}}\).

\subsection{\(\rm{H_2}\) and temperature}

The behavior of the chemical network under the radiation by cosmic ray protons is not well known. 
What we expect from the coupling between the primordial chemical network and cosmic rays is an enhancement in the number density of \(\rm{H_2}\), as cosmic rays ionize the atomic species and provide additional free electrons. This would eventually result in a net cooling of the gas by \(\rm{H_2}\) line emission if:
\begin{enumerate}
\item \(\rm{H_2}\) is produced on sufficiently short timescales and \(\rm{H_2}\) line cooling always outweighs cosmic ray heating;
\item  the gas is ionized only moderately to provide sufficient atomic hydrogen allowing for high \(\rm{H_2}\) formation rates. 
\end{enumerate}
The line cooling function \(\Lambda_{\rm{H_2}}\) of molecular hydrogen depends on the gas temperature \(T\). As in a low temperature gas the colliding particles on average do not possess enough kinetic energy to excite the rotational modes of molecular hydrogen, cooling by \(\rm{H_2}\) line emissions becomes more and more inefficient at lower temperatures. Therefore we expect a minimum temperature at which no further cooling by molecular hydrogen is possible and cooling is balanced by the heating of cosmic rays.

As \(\Lambda_{\rm{H_2}}\) depends on \(n_{\rm{H_2}}\), sufficient amounts of molecular hydrogen must be produced  on short timescales to radiate away the heat input to the medium. For the formation of H$_2$, free electrons and atomic hydrogen are needed. As cosmic rays ionize the medium they produce additional free electrons, but at the same time reduce the number of neutral hydrogen atoms. Therefore we expect a lower \(\rm{H_2}\) formation rate once a considerable fraction of the atomic hydrogen is ionized.

To assess all these effects we run simulations with gas of four different total number densities (\(1\)-\(1000\unit{cm^{-3}}\)) and cosmic ray energy injection rates \(\dot{\epsilon}_{CR}\) ranging from \(10^{-36}\) to \(10^{-24}\unit{erg\,cm^{-3}\,s^{-1}}\), with different SUV background fluxes \(F\). The evolution of the chemical network is followed over a time of \(0.47\unit{Gyr}\) during which \(\dot{\epsilon}_{CR}\) is assumed to be constant. After this integration time the gas has reached nearly thermal and chemical equilibrium, and not much change is observed beyond it.

The final values for the temperature \(T\) and the number density of molecular hydrogen \(n_{\rm{H_2}}\) are plotted in figure \ref{fig:Temp_300K} and figure \ref{fig:Temp_3000K} as a function of \(\dot{\epsilon}_{CR}\) for simulations with initial temperature \(T_0=300\unit{K}\) and \(3000\unit{K}\) respectively.

\begin{figure*}
	\centering
		\rotatebox{0}
	{
		\includegraphics[bb = 50 0 600 755,width=0.95\textwidth,clip=true]{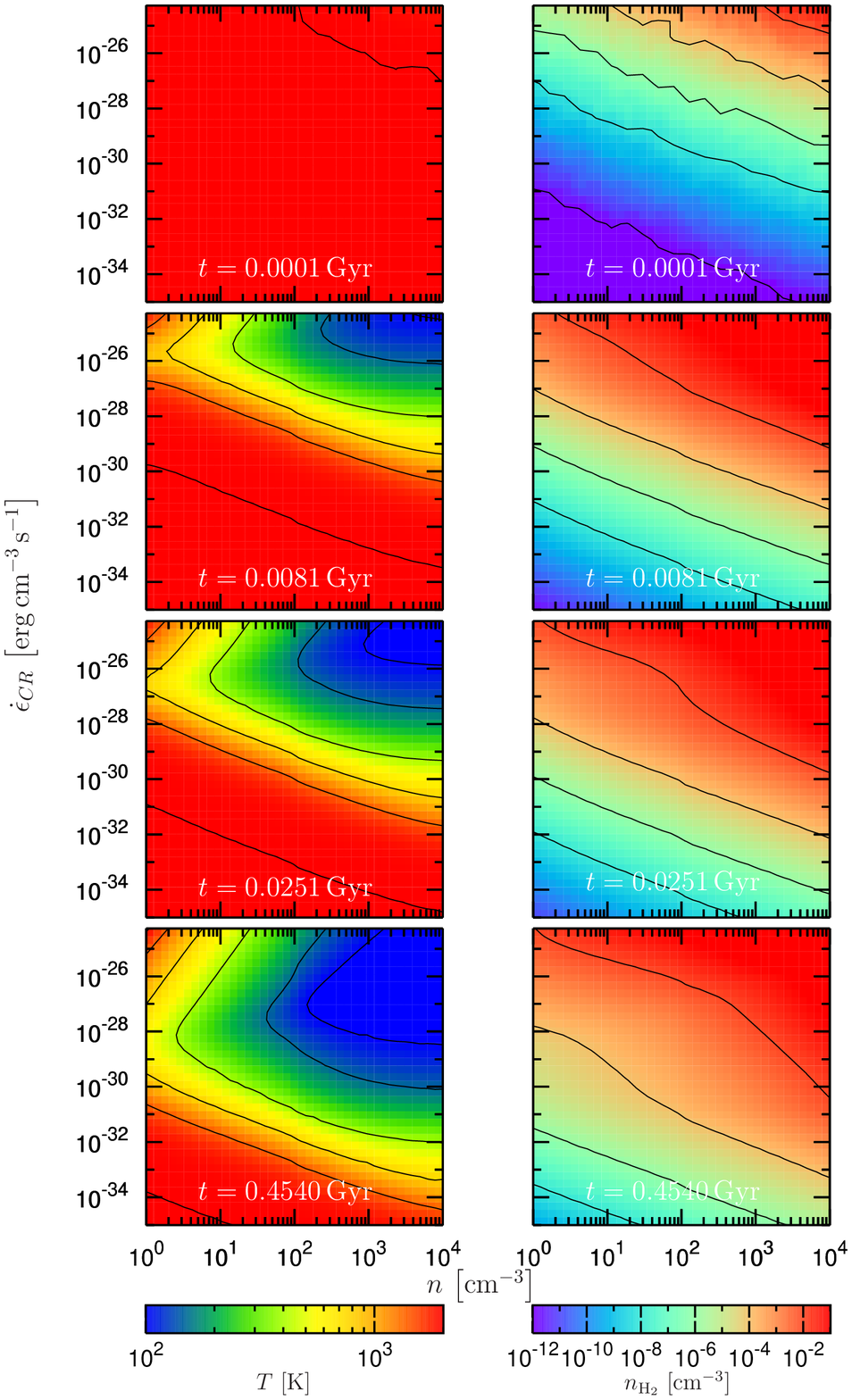}
	}
	\caption{Time evolution of the \(\dot{\epsilon}_{CR}\times n \)-plane for calculations with a neutral atomic gas at initial temperature \(T_0=2000\unit{K}\), embedded in a dissociating flux with \(F=10^{-21}\unit{erg\,cm^{-2}\,Hz^{-1}\,s^{-1}}\). The panels refer to different times. The left panels display the temperature evolution while the right panels show the corresponding values for \(n_{\rm{H_2}}\). The superimposed contour plots join points of same temperature or same \(\rm{H_2}\) number density.} 
	\label{fig:time2000a}
\end{figure*}

The overall observed effect is counterintuitive, as the gas cools with higher energy injection by cosmic rays. 
As can be seen in the plots, the temperature behavior (left panels) is highly dependent on the given environmental conditions, i.e. the total number density \(n\), the initial gas temperature \(T_0\), the cosmic ray source strength \(\dot{\epsilon}_{CR}\) and the SUV background flux. As expected from the previous discussions the final temperature is no monotonic function of the cosmic ray energy density injection rate \(\dot{\epsilon}_{CR}\). Beyond a certain value of \(\dot{\epsilon}_{CR}\), which also depends on the total gas number density \(n\), cosmic rays start heating the gas again instead of inducing further cooling.
In addition we observe that gas with lower density is more likely to be heated by cosmic rays than gas of higher density, as here production of molecular hydrogen is easier and \(\rm{H_2}\)-line cooling is more efficient.

The data also show a considerable impact of the SUV background on the thermal evolution of the medium.
A strong dissociating background suppresses the formation of molecular hydrogen and thus \(\rm{H_2}\) line cooling counterbalances the cosmic ray heating less efficiently. 
With a dissociating flux of \(F=10^{-18}\unit{erg\,cm^{-2}\,s^{-1}\,Hz^{-1}}\) the production of molecular hydrogen is suppressed by a factor of \(10^{-3}\) compared to the fiducial calculations with \(F=10^{-21}\unit{erg\,cm^{-2}\,s^{-1}\,Hz^{-1}}\), resulting in effectively stronger heating by cosmic rays.

The choice of the initial temperature is critical for the evolution of the gas, as can be seen by a comparison between figure \ref{fig:Temp_300K} and figure \ref{fig:Temp_3000K}, which immediately shows that the effect of cosmic ray induced cooling is much more efficient at high initial temperatures. This is due to several effects, the most important being the increase of the \(\rm{H^-}\) formation rate with temperature and the fact that \(\rm{H_2}\)-line cooling is more efficient at higher temperatures. Beside these effects the slight decrease of the recombination rates of the species \(\rm{H}\), \(\rm{D}\) and \(\rm{He}\) with temperature helps to increase the population of free electrons needed to catalyze the formation of \(\rm{H_2}\). 

It is interesting to note that with increasing initial temperature
\(T_0\) cosmic rays of lower intensity become more and more
important. While, for example, for a gas with \(T_0=300\unit{K}\) and
\(n=10\unit{cm^{-3}}\) the first observable cooling effect begins at
\(\dot{\epsilon}_{CR}\sim 10^{-30}\unit{erg\,cm^{-3}\,s^{-1}}\), the
cooling effect starts already at \(\dot{\epsilon}_{CR}\sim
10^{-33}\unit{erg\,cm^{-3}\,s^{-1}}\)  for the same gas simulated with
\(T_0=3000\unit{K}\).
This suggests that in the latter case cooling induced by cosmic rays
might be efficient also at number densities $<1$~cm$^{-3}$.

A closer comparison of figure \ref{fig:Temp_300K} and figure \ref{fig:Temp_3000K} reveals another interesting feature. 
At the lowest energy injection rates the production of molecular hydrogen in gas \(T_0=3000\unit{K}\) is increased by nearly one order of magnitude compared to the calculations with gas starting from \(T_0=300\unit{K}\). In contrast, at the highest simulated values of \(\dot{\epsilon}_{CR}\) the results for temperature \(T\) and \(n_{\rm{H_2}}\) are identical for both initial temperatures. In particular the results of calculations with these two initial temperatures are identical for values of \(\dot{\epsilon}_{CR}\)  higher than the energy injection rate at the minima in the temperature curves. Therefore beyond this point the chemical and thermal evolution of the gas does not depend on the initial conditions any more and all gas starting from whatever initial condition is converted to the same final state. Thus the minimum in the temperature curve defines a characteristic point for each gas.

\subsection{Time evolution of parameter space}

So far we have discussed the behavior of primordial gas under the influence of cosmic rays comparing only the final results of several calculations. It is instructive to study also the temporal evolution of the gas. Therefore, we display the evolution of the  \(\dot{\epsilon}_{CR}\times n \)-plane in time as a sequence of images. To have a satisfactory resolution of the parameter space we have run calculations for 40 different values of \(n\), in the range \(1\)-\(10^4\unit{cm^{-3}}\), and of \(\dot{\epsilon}_{CR}\) in the range \(10^{-36}\)-\(10^{-24}\unit{erg\,cm^{-3}\,s^{-1}}\). Each image of the \(\dot{\epsilon}_{CR}\times n \)-plane then consists of 1600 points, where each point represents a different simulation.

In figure \ref{fig:time2000a} we present some snapshots of a simulation with \(T_0=2000\unit{K}\) and \(F=10^{-21}\unit{erg\,cm^{-2}\,s^{-1}\,Hz^{-1}}\). It can be seen that molecular hydrogen forms from the very beginning of the simulation, at times when no change in temperature can be observed yet. Initially we observe the highest effect in correspondence with high values of the energy density injection rate \(\dot{\epsilon}_{CR}\) and total gas number density \(n\), while the effect for low values of \(\dot{\epsilon}_{CR}\) and \(n\) takes time to build up.

\section{Discussion}
\label{Discussion}

The calculations presented in this work give us a better understanding of the behavior of a primordial gas exposed to radiation of cosmic rays. In particular the parameter space maps, presented in figure \ref{fig:time2000a}, can be used to identify regions of interest in the early universe. Here we want to discuss these results in more detail.

The most interesting conclusion that can be drawn from our
calculations is that cosmic rays do not necessarily heat a gas, quite
the contrary they might contribute to cooling. However, this trend is
non-monotonic, as once the cosmic ray flux becomes larger than a critical
value that depends on the gas initial conditions, heating dominates. This
indicates that, to estimate the correct behavior of the thermal
evolution of a primordial gas under the influence of cosmic rays, 
\(\rm{H_2}\) chemistry should be included self-consistently.

We studied the response of primordial gas to cosmic ray radiation in a
variety of different environmental settings. In particular we were
interested in the influence of the total gas number density \(n\), the
initial temperature \(T_0\), the SUV background and the cosmic ray
intensity represented by \(\dot{\epsilon}_{CR}\). 
We observed that the effect of cosmic rays depends strongly on the
environment, especially on the initial temperature and
the strength of the dissociating photon background. The results
presented here therefore enable us to clearly define the environmental
conditions for which this effect might be of interest. The following
conditions are necessary for the cooling induced by cosmic rays to be
efficient (although some cooling can be observed also for lower
temperatures and densities):
\begin{enumerate}
\label{enu:cool_cond}
\item A dissociating photon flux with \(F<10^{-18}\unit{erg\,cm^{-2}\,Hz^{-1}\,s^{-1}}\).
\item Initial temperature of the gas \(\sim10^{3}\unit{K}\). 
\item Total number densities of the gas \(n\ge1\unit{cm^{-3}}\).
\item Cosmic ray sources with \(\dot{\epsilon}_{CR}>10^{-33}\unit{erg\,cm^{-3}\,s^{-1}}\).
\end{enumerate}
We can therefore think of a variety of possible astrophysical scenarios in which cosmic rays might efficiently influence the chemistry of a primordial gas.
For example, the first two conditions are in excellent agreement with the environment that can be found at the epoch of first structure formation, when objects with \(T_{vir}\le10^4{\unit{K}}\) collapse. As the cosmic microwave background is already much too weak to ionize, the only possible source for ionizing photon radiation would be stars that have not formed in sufficient amounts at these times. During the process of structure formation shock heated gas can contribute to the ultraviolet background by thermal emissions. However, these fluxes are expected to be small, especially for small structures, and in case of the SUV background do not exceed values of \(F\sim10^{-21}\unit{erg\,cm^{-2}\,Hz^{-1}\,s^{-1}}\) \citep{Miniati}. In addition, X-rays can contribute to ionization. These high energetic photons arise from free-free interactions of the electrons in the shock heated gas. As soft X-rays do not penetrate deeply into clouds and the flux of hard X-rays is small, the ionization is expected to be dominated by cosmic rays in the regimes considered in this work \citep{book_interstellar_medium}. It should be noted that, to have a proper estimate of the relative importance of different ionization sources, a self-consistent calculation should be done.
A more thorough study on effects of CR on early structure formation
is deferred to future work.

The main goal of the calculations presented above was to get a better understanding of the cosmic ray chemistry and effects which affect the \(\rm{H_2}\) formation. Therefore in all calculations no volume evolution was considered. This introduces some systematic errors when we want to apply the results to real cosmological situations.
As the gas expands or contracts when it is heated or cooled, the total
number density of the gas particles changes as a response to
temperature change. Therefore, when the gas starts to cool, its number
density increases leading to a higher formation rate of
\(\rm{H_2}\). This effect might increase the overall \(\rm{H_2}\)
formation rate for a gas which experiences cooling induced by cosmic
rays. To properly assess the impact of the volume evolution on our
calculations, a detailed treatment of the gas collapse and the effects
of cosmic ray pressure should be included. This is beyond the scope of
this study and we defer a more thorough investigation to future work.

The conclusions drawn in this paper rely on the presence of CR protons in the early universe, but our knowledge about primordial CR is very limited. As already mentioned in the Introduction some possible sources of primordial CR are decaying primordial black holes, structure formation shocks or the very first Supernovae.
There is also the possibility that weakly interacting massive
particles (WIMPS) decay into protons and anti-protons
\citep{Diehl}. This would be an interesting source for cosmic ray
protons as it is homogeneously distributed in space. Unfortunately the
proton and antiproton fluxes arising from the decaying WIMPS are
expected to be too low to have any interesting impact on the gas
cooling \citep{Diehl}.

A more promising cosmic ray source in the absence of stars may be
structure formation shocks. Cosmological shock waves form abundantly
in the course of structure formation, both due to infalling pristine
cosmic gas, which accretes onto filaments, sheets and halos, as well
as due to supersonic flows associated with merging structures
\citep{Pfrommer}. These shocks are able to dissipate gravitational
energy associated with hierarchical clustering into thermal energy of
the gas contained in dark matter halos. Beside this, shocks are able
to accelerate ions of the high energy tail of the Maxwellian velocity
distribution of a thermal medium by diffusive shock acceleration
\citep{Pfrommer}. This acceleration process produces a cosmic ray
population with a power-law distribution of the particle momenta. For
this reason structure formation shocks seem to be plausible cosmic ray
sources. As in addition we do not expect strong UV fluxes during early
structure formation \citep{Miniati} our calculations could be applied
to such a configuration. Numerical studies suggest that the mean
energy density injection rate of cosmic rays from structure formation
shocks at \(z=10\) is
\(\dot{\epsilon}_{CR}\approx10^{-33}\unit{erg\,cm^{-3}\,s^{-1}}\),
where this value is averaged over a co-moving volume of \(\approx
(143\unit{Mpc})^3\) \citep{Pfrommer}. Thus, the local value of
\(\dot{\epsilon}_{CR}\) might be much higher.

Once the very first stars appear, much more powerful cosmic ray
sources become available. These early massive stars (\(M\sim
100 {\rm M}_{\sun}\) ) form from metal-free gas through molecular hydrogen
cooling \citep{Abel_bryan} and are likely to end either as supernova
or a black hole \citep{Abel_wise}. Supernovae are believed to be very
efficient particle accelerator \citep{Biermann_introduction} and
therefore could easily increase the abundance of an early cosmic ray
population. This could help to make the effect observed in our
calculations more efficient and trigger the formation of second
generation stars. 
Unfortunately primordial stars produce a lot of UV radiation
which might severely hamper the production of molecular hydrogen
\citep{Abel_wise}. Also, as already pointed out earlier, in the
presence of ionizing photons the effect of cosmic ray ionization is
usually sub-dominant, and if the medium is partly ionized cosmic rays 
will rather heat by Coulomb heating than induce
\(\rm{H_2}\)-line cooling.
One should also note that stars produce metals which are far more efficient coolants than molecular hydrogen. Therefore, in order to observe the effects described in this work, cosmic rays have to extend further into space than the pollution by metals.
For these reasons we do not expect any significant effect of cosmic ray induced cooling in the vicinity of the sites of the very first stars and supernovae.

If, however, cosmic rays would be able to travel further away than the ionization front and the metal polluted area built up by the progenitor, they might contribute to the formation of molecular hydrogen in the ambient neutral medium.
Anyway, such a scenario would require detailed knowledge about the propagation properties of cosmic rays and magnetic fields in the early universe, which we presently do not have. 

\section{Conclusion}
\label{Conclusion}

In this work we studied the influence of an early cosmic ray
population on the chemistry of primordial gas. We developed a
computer code to couple self-consistently the chemistry of a gas of
primordial composition with the spectral evolution of cosmic rays
permeating this medium. The code was then used to study the response
of the gas in a variety of environmental
settings. The main results of these calculations can be summarized as
follows:

\begin{enumerate}
\item{Cosmic rays do not necessarily heat, quite the contrary they can contribute efficiently to the cooling of a medium.}
\item Cosmic rays influence the chemical state of a gas appreciably. In particular they catalyze the formation of molecular hydrogen.
\item The impact of cosmic rays on the gas chemistry depends highly on the environmental conditions.
\end{enumerate}

However, the detailed processes of cosmic ray production and propagation in the early universe are yet unclear and further investigation will be needed to completely understand a potential relevance of cosmic rays for early structure formation.

\section*{Acknowledgments}
This paper has benefited from a discussion with Christoph Pfrommer who supported us with reasonable limits on the CR energy injection rates from early structure formation shocks. Further we thank Andre Henri Waelkens and Mona Frommert for many helpful discussions and support in the course of this work. I also like to thank Daniele Galli and Francesco Palla who provided us with the numerical Data of their work and allowed us to reuse them for our purpose.

\bibliography{paper}
\bibliographystyle{mn2e}
%%% Local Variables: 
%%% mode: latex
%%% TeX-master: t
%%% End: 

\appendix
\section{Heating and cooling rates}
\label{coolrat}
The cooling rates used in this work are given in table \ref{tab:cool_reactions} together with the corresponding references.

The kinetic energy of particles (i.e. electrons) produced by photo-ionization and photo-dissociation will be the heat sources. According to \cite{shapiro_kang} the heating rate due to these reactions is given by:

\begin{equation}
\label{eq:phot_heat}
\Gamma= n_i \int_{\nu_{th}}^{\infty}\left(h\nu -h\nu^i_{th}\right)\,F(\nu)\, \sigma^i(\nu) \frac{d\nu}{h\nu}\, ,
\end{equation}
where \(n_i\) and \(\nu^i_{th}\) are the density and threshold energy of species \(i\), respectively, \(F(\nu)\) and \(\sigma^i(\nu)\) are the radiation flux and the cross section for the reaction, respectively. Note that no radiative transfer is considered in equation \ref{eq:phot_heat}. 

\begin{table}
	\centerline{
	\begin{tabular}{llll}
	\hline
	Cooling rate & Reference \\
	\hline
	\(\Lambda _{line}(\rm{H})\)		& \cite{Cen} and \cite{Black}	\\
	\(\Lambda _{line}(\rm{He})\)	& \cite{Cen} and \cite{Black}	\\
	\(\Lambda _{line}(\rm{He^+})\)	& \cite{Cen} and \cite{Black}	\\
        \(\Lambda _{ion}(\rm{H})\)&\cite{Cen} and \cite{shapiro_kang}\\
	\(\Lambda _{ion}(\rm{He})\)&\cite{Cen} and \cite{shapiro_kang}\\
	\(\Lambda _{ion}(\rm{He^+})\)&\cite{Cen} and \cite{shapiro_kang}\\
	\(\Lambda _{ion}(\rm{He})\)&\cite{Cen} and \cite{shapiro_kang}\\
	\(\Lambda _{rec}(\rm{H^+})\)&\cite{Black} and \cite{Spitzer_book}\\
	\(\Lambda _{rec}(\rm{He^+})\)&\cite{Black} and \cite{Spitzer_book}\\
	\(\Lambda^{de}_{rec}(\rm{He^+})\)&\cite{Black} and \cite{Spitzer_book}\\
	\(\Lambda _{rec}(\rm{He^{++}})\)&\cite{Black} and \cite{Spitzer_book}\\
	\(\Lambda _{Brems}\)&\cite{Black}\\
	\(\Lambda _{compton}\)&\cite{peebles}\\
	\(\Lambda _{\rm{H^-}\mbox{diss}}\)&\cite{shapiro_kang}\\
	\(\Lambda_{\rm{H_2}\, \mbox{formation}}\)&\cite{Abel_chem}\\
	\(\Lambda_{\rm{H_2}\, \mbox{diss}}\)& \cite{shapiro_kang}\\
	\(\Lambda_{\rm{H_2}\, \mbox{line}}\)&\cite{Galli_Palla}\\
	\hline
	\end{tabular}}
	\caption{Cooling rates and the corresponding references.}
	\label{tab:cool_reactions}
	\end{table}

\section{Chemical rates}
\label{chemrat}
The chemical reactions implemented in our code are given in table \ref{tab:reactions} together with the references to the chemical rates.
\begin{table}
	\centerline{
	\begin{tabular}{llll}
	\hline
	Reaction & Reference \\
	\hline
	\reack{k1}& \cite{janev_etal}\\
	\reack{k2}& \cite{Abel_chem}\\
	\reack{k3}& \cite{janev_etal}\\
	\reack{k4}& \cite{Cen_1992} and \cite{Aldrovandi}\\
	\reack{k5}&  AMDIS Database (1989) see \cite{Abel_chem}\\
	\reack{k6}& scaling (e.g. see \cite{Osterbrock})\\
	\reack{k7}&    \cite{Abel_chem}\\
	\reack{k8}&    \cite{Abel_chem}, \cite{shapiro_kang}\\
	\reack{k9}&    \cite{shapiro_kang}\\
	\reack{k10}&    \cite{karpas_etal}\\
	\reack{k11}&    \cite{Abel_chem}\\
	\reack{k12}&    \cite{donahue_shull}\\
	\reack{k13}&    \cite{dove_mandy}\\
	\reack{k14}&    \cite{janev_etal}\\
	\reack{k15}&    \cite{Abel_chem}\\
	\reack{k16}&    \cite{dalgarno_lepp}\\
	\reack{k17}&    \cite{Abel_chem}\\
	\reack{k18}&    \cite{Abel_chem}\\
	\reack{k19}&    \cite{dalgarno_lepp}\\
	\reack{k29}& \cite{Galli_Palla}\\
	\reack{k31}&   \cite{Galli_Palla}\\
	\reack{k32}&    \cite{Galli_Palla}\\
	\reack{k33}&    \cite{Smith_etal}\\
	\reack{k34}&    \cite{Smith_etal}\\
	\reack{k50}&    \cite{Roberge}\\
	\reack{k51}&    \cite{karpas_etal}\\
	\reack{k52}&   \cite{Yousif}\\
	\(\rm{H}+\gamma \to \rm{H^+}+\rm{e^-}\)&       \cite{Osterbrock_1974}\\
	\(\rm{He}+\gamma \to \rm{He^+}+\rm{e^-}\)&       \cite{Osterbrock_1974}\\
	\(\rm{He^+}+\gamma \to \rm{He^{++}}+\rm{e^-}\)&      \cite{Osterbrock_1974}\\
	\(\rm{H^-}+\gamma \to \rm{H}+\rm{e^-}\)  &     \cite{Dejong}\\
	\(\rm{H_2}+\gamma \to \rm{H^+_2}+\rm{e^-}\) &      \cite{oneil_reinhardt}\\
	\(\rm{H^+_2}+\gamma \to \rm{H}+\rm{H^+}\)    &   \cite{Osterbrock_1974}\\
	\(\rm{H^+_2}+\gamma \to 2\rm{H^+}+\rm{e^-}\) &      \cite{shapiro_kang}\\
	\(\rm{H_2}+\gamma \to\rm{H^*_2}\to \rm{H} +\rm{H}\)&       \cite{Abel_chem}\\
	\(\rm{H_2}+\gamma \to\rm{H} +\rm{H}\)   &    \cite{Abel_chem}\\
	\(\rm{H}+\gamma_{CMB} \to \rm{H^+}+\rm{e^-}\)&       \cite{Galli_Palla}\\
	\(\rm{He}+\gamma_{CMB} \to \rm{He^+}+\rm{e^-}\)&     \cite{Galli_Palla}\\
	\(\rm{He^+}+\gamma_{CMB} \to \rm{He^{++}}+\rm{e^-}\)&      \cite{Galli_Palla}\\
	\(\rm{H^-}+\gamma_{CMB} \to \rm{H}+\rm{e^-}\)  &     \cite{Galli_Palla}\\
	\(\rm{H_2}+\gamma_{CMB} \to \rm{H^+_2}+\rm{e^-}\) &      \cite{Galli_Palla}\\
	\(\rm{H^+_2}+\gamma_{CMB} \to \rm{H}+\rm{H^+}\)    &   \cite{Galli_Palla}\\
	\(\rm{H^+_2}+\gamma_{CMB} \to 2\rm{H^+}+\rm{e^-}\) &      \cite{Galli_Palla}\\
	\(\rm{D}+\gamma_{CMB} \to\rm{D^+}+\rm{e^-}\)&       \cite{Galli_Palla}\\
	\(\rm{HeH^+}+\gamma_{CMB} \to\rm{He} +\rm{H^+}\)   &    \cite{Galli_Palla}\\
	\hline
	\end{tabular}}
	\caption{Reactions and references for the different chemical rates.}
	\label{tab:reactions}
	\end{table}

\bsp

\label{lastpage}

\end{document}